\documentclass[ms]{aastex}
\usepackage{spr-astr-addons}
\usepackage{url}

\RequirePackage{color}

\def\arcsec{\hbox{$^{\prime\prime}$}}

\newcommand{\emailb}{wilhelm@mps.mpg.de}
\newcommand{\emaila}{bholadwivedi@gmail.com}

\begin{document}
  \title{Solar coronal plumes and the fast solar wind}

  \author{Bhola N. Dwivedi}
  \affil{Department of Physics, Indian Institute of Technology
(Banaras Hindu University), Varanasi-221005, India \\ \emaila}
  \and
  \author{Klaus Wilhelm}
  \affil{Max-Planck-Institut f\"ur Son\-nen\-sy\-stem\-for\-schung (MPS),
    37077 G\"ottingen, Germany \\ \emailb}

  \abstract
The spectral profiles of the coronal Ne\,{\sc viii} line at 77~nm have
different shapes in quiet-Sun regions and coronal holes (CHs). A single
Gaussian fit of the line profile provides an adequate approximation in
quiet-Sun areas, whereas a strong shoulder on the long-wavelength side is a
systematic feature in CHs. Although this has been noticed since 1999, no
physical reason for the peculiar shape could be given. In an attempt to
identify the cause of this peculiarity, we address three problems that
could not be conclusively resolved in a review article by a study team of the
International Space Science Institute \citep[ISSI;][]{Wiletal}:
(1) The physical processes operating at the base and inside of
plumes as well as their interaction with the solar wind (SW).
(2) The possible contribution of plume plasma to the fast SW streams.
(3) The signature of the first-ionization potential (FIP) effect between
plumes and inter-plume regions (IPRs).
Before the spectroscopic peculiarities in IPRs and plumes in polar
coronal holes (PCHs) can be further investigated with the instrument
Solar Ultraviolet Measurements of Emitted Radiation (SUMER) aboard the Solar
and Heliospheric Observatory (SOHO), it is mandatory to summarize the
results of the review to place the spectroscopic observations into context.
Finally, a plume model is proposed that satisfactorily explains the
plasma flows up and down the plume field lines and leads to the shape of
the neon line in PCHs.


  \maketitle

\section{Introduction}
\label{s.introd}

\begin{figure*}[t]
\centering
\includegraphics[width=\textwidth]{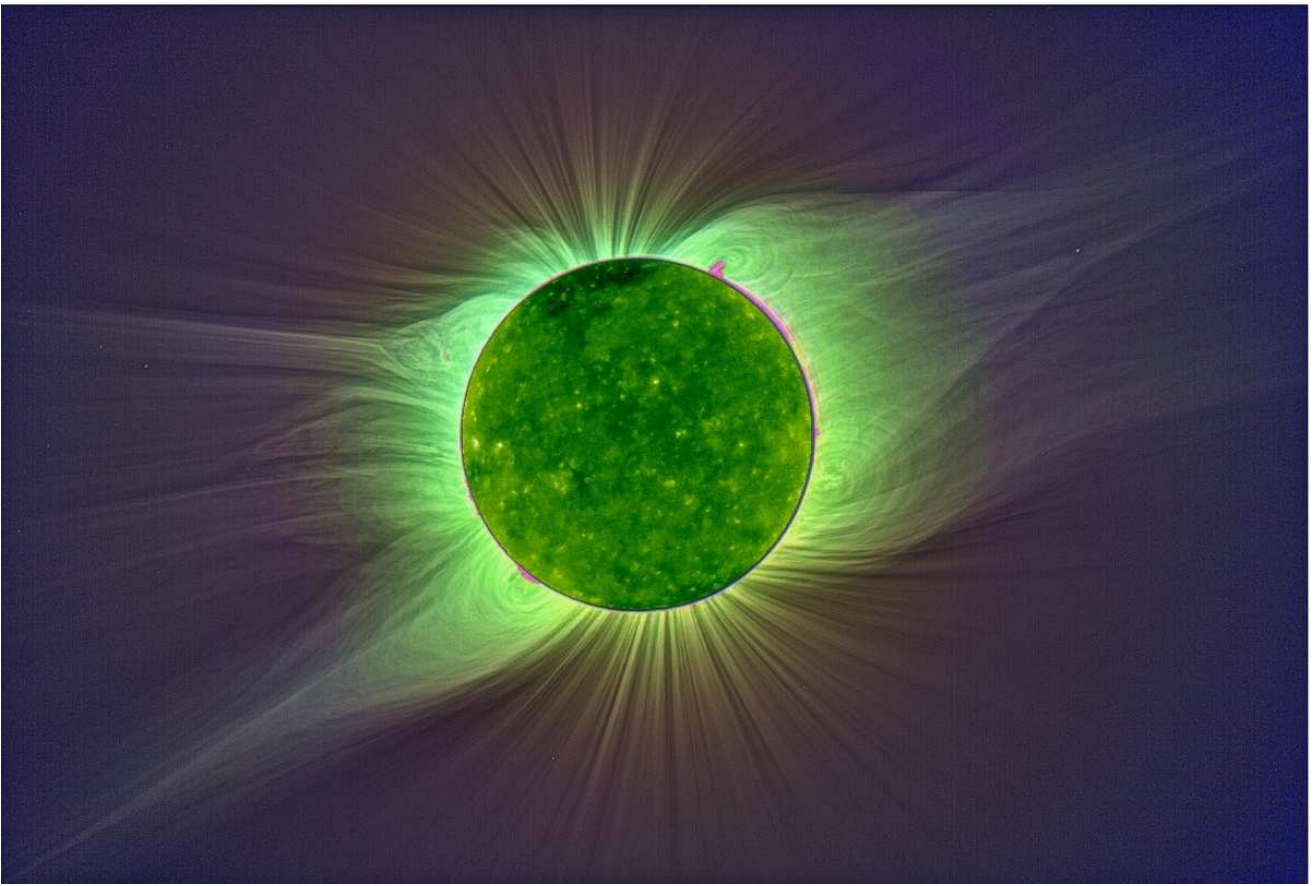}
\caption{\label{Figcor} The solar corona during the total eclipse
on 1 August 2008 observed from
Mongolia. The corona at solar minimum conditions has wide PCHs
with reduced radiation, open magnetic field lines and
many plume structures. At lower latitudes
closed field-line regions dominate the corona and extend into coronal streamers
\citep[from][composite eclipse image by M. Druckm\"uller, P. Aniol and V.
Ru\v{s}in]{Pasetal}. An image in 19.5~nm of the solar disk taken from
the Extreme ultraviolet Imaging Telescope (EIT) \citep{Boudine} on
SOHO at the time of the eclipse
has been inserted into the shadow of the Moon \citep{Wiletal}.}
\end{figure*}

In a recent review \citep{Wiletal}, many aspects of the solar coronal
plume phenomenon have been presented.
In most cases the authors of the review, members of a study team of the
International Space Science Institute (ISSI), Bern, arrived at conclusive
results. However, some open points remained, of which we consider
three. They have been formulated with the abbreviations
IPR (inter-plume region), SW (solar wind), and FIP
(first-ionization potential):
\begin{enumerate}
\item [-] Although models of plumes and their formation are available,
an exact
description of the physical processes operating at the base and inside of
plumes as well as their interaction with the SW is still outstanding.
\item [-] Is there any contribution of plume plasma to the fast SW streams at
all?
\item [-] What produces the clear FIP effect signature between plumes and
IPRs?
\end{enumerate}
In this paper, we propose\,--\,based on observational
data\,--\,tentative solutions in these problem areas.

Earlier review papers on plumes have been published as well,
and we refer the reader to them for a general introduction to this solar
phenomenon \citep[e.g.,][]{Hul50,Sai65,NewHar,DeFetal}. Nevertheless,
it is necessary to list some of the basic properties of plumes and IPR as
they are known at present from observations described by \citet{Wiletal}
and references therein:\\
Polar plumes delineate magnetic field lines of the minimum corona in PCHs and
expand super-radially in the low
$\beta$-regime of the corona as can be seen from Fig.~\ref{Figcor}.
Plumes observed in WL (white light) and
VUV (vacuum ultraviolet) result from plasma density enhancements in CHs.
The electron density ratio between plumes and IPRs is between three and
seven in the low corona and decreases at greater heights.
The electron temperature in plumes is $T_{\rm e} \leq 1$~MK. In IPRs it is
higher by $\approx$\,0.2~MK with a tendency of even higher values at greater
heights. A plume shows some evolution during its lifetime.
Footpoints of beam plumes lie near magnetic flux concentrations
interacting with small magnetic dipoles. The reconnection activity generates
heat near the base of a plume and leads to jets that probably provide some of
the plume plasma. The SW outflow velocity is higher in IPRs than in plumes.
Plumes and IPRs have a distinctly different abundance composition, in the
sense that the ratio of low-FIP/high-FIP elements is much larger in plumes
than in IPRs.
Rosettes in the chromospheric network could be of importance for the plume
formation.

The following definition has been used for $\beta$:
%
\begin{equation}
\frac{p}{p_{\rm mag}} = \frac {n\,k_{\rm B}\,T}{(B^2/2\mu_0)} = \beta~,
\label{beta}
\end{equation}
where $p$ is the plasma pressure and $p_{\rm mag}$ is the magnetic pressure
with $n$ the particle density, $k_{\rm B}$ the Boltzmann constant, $T$ the
plasma temperature, $B$ the magnetic flux density, and $\mu_0$ the vacuum
permeability.

The plasma of PCHs is optically thin for VUV lines. Nevertheless, it was
possible to identify two different plasma regimes, plumes and IPR,
by studying density- and temperature-sensitive line ratios \citep{Wil06}.
From EIT observations, \citet{Gab03} have found that plumes must occur
with two different morphologies, beam plumes and curtain or network plumes,
because the latter appear to be aligned along network lanes. Network plumes, and
probably beam plumes, are composed of individual micro-plumes \citep{Gab09}.
This aspect was further studied by \citet{Patetal} who found
typical beam plumes with a localized cross-section and those with an
elongated cross-section as expected for network plumes. Their tomography
results show that intermediate configurations also exist.

Fig.~\ref{Figcor} clearly shows that the field lines of a PCH open into
interplanetary space. The observations of the spacecraft Ulysses demonstrate
that on these field lines the fast solar wind escapes from the Sun with
asymptotic speeds of approximately 800~km\,s$^{-1}$
\citep[cf., e.g.,][]{Wocetal,McCetal}.

\section{The neon emission line near 77~nm}
\label{s.neon}

The transition $2{\rm s}\,^2{\rm S}_{1/2}-2{\rm p}\,^2{\rm P}_{3/2}$ in the
Ne$^{7+}$ ion leads to a prominent solar emission line in the Ne\,{\sc viii}
spectrum near 77~nm. The spectroscopic observations of this line
are of major importance in this study and therefore some background
information might be helpful.
The first wavelength determination
$\lambda_0 = 77.042$~nm
with a standard uncertainty of 0.003~nm was performed in the laboratory by
\citet{Fawetal}.
The large Doppler width of the line emitted from
high-temperature plasmas limits the accuracy of such measurements. Solar
observations, therefore, provide the best values of the rest wavelength in
vacuum $\lambda_0 = (77.0428 \pm 0.0003)$~nm \citep{Dametal}.
The
Ne\,{\sc viii} line is formed at an electron temperature of 620\,000~K
\citep[cf.,][]{Wil02}.
The contribution function has a long tail towards
higher temperatures typical for lithium-like ions. The line is thus
ideally suited for studies of the upper transition region and its interface
with the {\em low corona}. Measurements of Doppler shifts of this line in
quiet-Sun (QS)
regions initially provided inconsistent results for the average shift. The
problems were, however, related to the inaccurate knowledge of the vacuum
rest wavelength \citep{Dosetal,Hasetal1,Breetal,Chaetal}.
Later studies showed an average blue shift of
$\approx 1$~km~s$^{-1}$ in QS regions
\citep{PetJud,Dametal},
and a more pronounced average blue shift of $\approx 6$~km~s$^{-1}$ in PCHs.


\section{Observed outflow speeds in polar coronal holes}
\label{s.outflow}

%
\begin{figure*}[t]
\centering
\includegraphics[width=\textwidth]{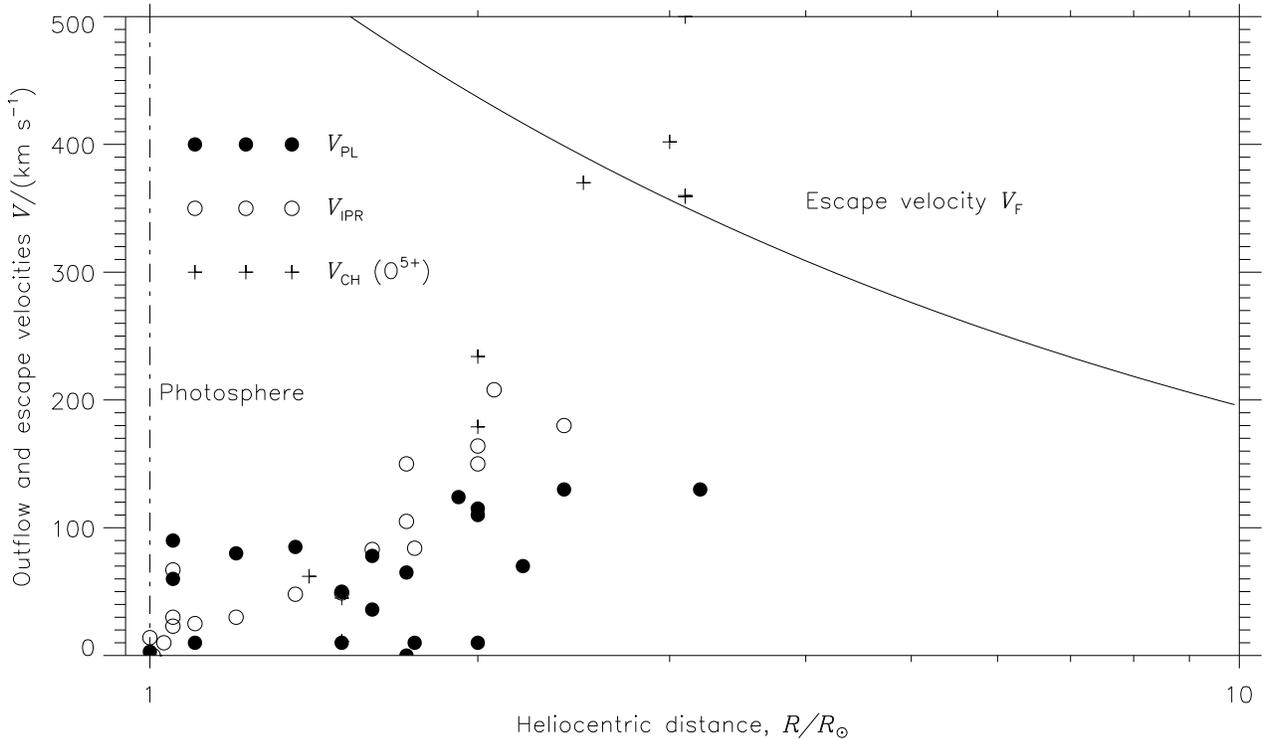}
\caption{\label{f.escape}
Flow speeds observed in polar coronal holes (PCHs). Plume and
inter-plume-region (IPR) measurements are plotted
separately. The observations in the O\,{\sc vi} line emitted by O$^{5+}$
ions refer to a PCH.
Also shown is the escape velocity as function of the heliocentric distance
\citep[cf.,][]{Wiletal}.}
\end{figure*}

Before an attempt can be undertaken to answer the question:
\emph{Are there plume signatures in the fast solar wind?} the observed
outflow speeds in plumes and IPR have to be considered as well as
the elemental composition of the solar photosphere and the polar corona.

Strong outflows were observed by \citet{Hasetal9} in a PCH above bright areas
as seen in the Si\,{\sc ii} 153.3~nm line at the intersections of chromospheric
network lanes. However, these areas are rather dark in the Ne\,{\sc viii}
radiance maps and have a typical flow component of Ne$^{7+}$ ions of
10~km\,s$^{-1}$ along the line of sight. The ion outflow speed can then be
obtained with the magnetic field model of \citet{Banetal} as 14~km\,s$^{-1}$.
No outflow is observed in BPs at the base of polar plumes \citep{Wil00}.

Funnel-shaped magnetic flux tubes from the photosphere to the corona
are typical features in PCHs \citep[cf., e.g.,][]{Gab76,Tuetal,Itoetal}.
These funnels are seen as source regions of the fast SW, but they can also
contain coronal plumes. One funnel analysed by \citet{Tuetal} in their
Figs.~1(F) and 4 at
$x = 50\arcsec$~and $y = 175\arcsec$~does not show any outflow speed.
It had earlier been identified as a plume \citep{Wil00}.
Two plumes seen by \citet{Hasetal9} also appear to be stationary.

Outflow speeds observed in PCHs by many researchers have been compiled
by \citet{Wiletal} and are included here in Fig.~\ref{f.escape} together with
the escape velocities at heliocentric distances, $R$:
%
\begin{equation}
V_{\rm F}(R) = \sqrt{\frac{2\,G_{\rm N}\,M_\odot}{R}} ~,
\label{Escape}
\end{equation}
where $G_{\rm N}$ is the gravitational constant and $M_\odot$
the mass of the Sun.
The speeds measured in IPRs and PCHs (without distinction between IPRs
and plumes, which have a small filling factor) increase with increasing
heliocentric distance without too much scatter and attain escape velocities
near $R = 3~R_\odot$, whereas the speeds published for plumes vary
considerably and nowhere reach the escape velocity. We argue that there are
probably two reasons for the high variability: (1) Plumes evolve during their
lifetime and may display different characteristics at different stages.
(2) Jets with high outflow speeds are often observed near the footpoints
of plumes \citep[e.g.,][]{Raoetal,Patetal,RaoSte}.

This is in line with a suggestion by \citet{Hul50} that plumes are rather
static with occasional plasma injections along the field lines. This has been
confirmed by the findings of \citet{Sheetal} that the direction of time is
easy to see with the Large Angle Spectroscopic Coronagraph (LASCO) on SOHO
by tracing lateral inhomogeneities in coronal streamers, but difficult to
identify over PCHs.


\section{First-Ionization Potential effects in the solar atmosphere
and the solar wind}
\label{s.FIP}

In the corona, the abundances of elements with respect to the photosphere
vary and the FIP effect plays a
dominant r$\hat{\rm o}$le. Elements with a FIP value of $I_{\rm X} < 10$~eV
are defined as low-FIP elements and those with FIP of $I_{\rm X} > 10$~eV as
high-FIP elements, separated by the photon energy $h\,\nu = 10$~eV of
the H\,{\sc i} Ly\,$\alpha$ line.

The photosphere is generally assumed to represent the elemental
composition of the outer layers of the Sun. This could be confirmed by
\citet{SheSol} who showed that
only a very minor part of the element segregation observed in
the outer solar atmosphere seems to take place in photospheric
and subphotospheric layers.
\citet{WidFel9} found FIP effects in strong plumes, whereas no
significant FIP effect was observed in an IPR of a CH
\citep[cf.,][]{Fel98,Lan08}.
According to \citet{Dosetal8}, the Si/Ne abundance ratio in IPRs
in CHs is close to the photospheric value at temperatures near $10^6$~K.
The abundance ratio of magnesium (a low-FIP element) to neon (a high-FIP
element) in plumes is enhanced relative to IPRs by factors of 1.5 and 3.5
\citep{WilBod,Youetal}.
In Fig.~25 of \citet{Wiletal},
plume and IPR observations are compiled to characterize the
changes in elemental abundances over a PCH. Plumes can clearly be identified
against the IPRs by their high electron densities obtained from the
density-sensitive Si\,{\sc viii} (144.6, 144.0)~nm ratio and the lower
electron temperatures evident in the line ratio of two ionization stages of
silicon. Both the low-FIP elements magnesium and sodium are enriched relative
to the high-FIP element neon in plumes with respect to IPRs.

\citet{WidFel2} have  found in active regions (AR) that the
confinement time of a plasma
is a decisive parameter for abundance variations.
A FIP bias of nearly ten was reached
after $\approx 6$~d. If these findings can be applied to plumes, we would
expect confinement times of a day or so\,--\,not too different
from plume and BP lifetimes of days \citep[e.g.,][]{Wan98,DeF01a,Wiletal}.

The different elemental compositions of plumes and IPRs thus suggest that
plumes in contrast to IPRs provide some kind of containment for the solar
plasma for a period of days, in which the FIP effect can operate.


\section{Are there plume signatures in the fast solar wind?}
\label{s.SW}

If the IPRs are indeed the source regions of the fast SW, no composition
changes would be expected in the high-speed streams in accordance with
observations of \citet{Geietal}. \citet{Hebetal} also reached the conclusion
that the SW originating in regions of open magnetic field, would
probably not contain matter with any significant mass
fractionation.

\citet{Thi90} identified with the
help of plasma and magnetic field data obtained by the two Helios solar probes
41 fast SW streams between 0.3~ua and 1~ua often with a strong
anticorrelation between the variations in the gas pressure and the
magnetic pressure were found while the total pressure was nearly constant.
Ulysses observations \citep{Reietal} of the high-latitude SW have shown that on
time scales of less than one day, the polar SW is dominated by pressure
balance structures (PBSs). Fluctuations of the plasma $\beta$ within PBSs
appear to be strongly correlated with fluctuations in the helium abundance.
The authors suggest an interpretation of the high $\beta$ portion of PBSs as
the SW extensions of polar plumes. However, the abundance of helium
(a high-FIP element) should not be enhanced in plumes, if the neon observations
are taken into account.

Direct observation of plumes with SOHO instruments
have been made up to $15~R_\odot$,
``where they fade into the background noise'' according to \citet{DeFetal}.
Very strong plumes
could be followed to $30~R_\odot$, but beyond that distance there is no clear
indication for the presence of plume plasma in the SW
\citep[see, e.g.,][]{Poletal,DeF01b,Wiletal}.

Microstreams\,--\,identified in Ulysses data\,--\,have been analysed by
\citet{Neuetal}, who concluded that these were not to be identified with plumes.
The same result was reported by \citet{Steetal}, because no significant
depletion of the Ne/Mg abundance and charge-state deviation in these structures
could be detected.


\section{Spectroscopic peculiarities in polar coronal holes}
\label{s.pecul}
%
\begin{figure*}[t]
\centering
\includegraphics[width=\textwidth]{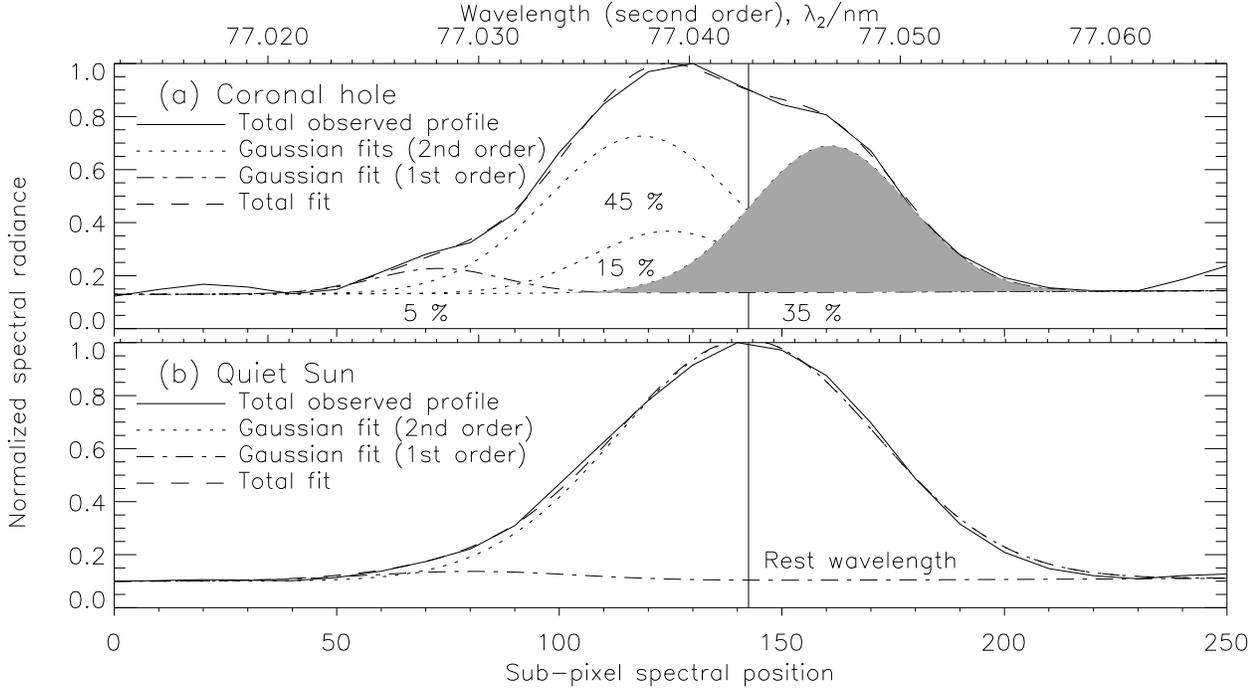}
\caption[]{\label{f.second_order} Spectral profiles covering the
Ne\,{\sc viii} line recorded with detector~A of SUMER in the second order
in a PCH on the solar disk {\bf (a)} and
in a QS region {\bf (b)}.
The profiles are normalized to one and have been approximated by three
Gaussian fits shown in dotted lines for second-order contributions and as
dashed-dotted line for a suspected first-order blend of 5~\% at
$2 \times 77.027$~nm.
The rest wavelength of
the Ne\,{\sc viii} line is indicated at $\lambda_0 = 77.0428$~nm.
The total profiles are consistent with a mean blueshift of $6.2$~km\,s$^{-1}$
in PCHs and $0.8$~km\,s$^{-1}$ in QS areas \citep{Dametal}. In panel~(a)
the 45~\% peak is shifted to the blue by $19$~km\,s$^{-1}$,
the 15~\% peak by $\approx 14$~km\,s$^{-1}$, and the 35~\% peak
by $15$~km\,s$^{-1}$ to the red side.}
\end{figure*}

In most of the studies performed with the SUMER instrument on SOHO
\citep{Wil95},
it was noticed that the profiles of the Ne\,{\sc viii} line were of
Gaussian shape in QS regions and in the corona above the limb, but
exhibited a strong shoulder on the long-wavelength wing if seen in
PCHs on the solar disk\footnote{All raw data acquired are in the public
domain and can be obtained either from the SOHO Archive or from the SUMER
Image Database at
{\tt www2.mps.mpg.de/\\projects/soho/sumer/FILE/SumerEntryPage.html} (accessed
on 16 Dec. 2014).}.
Attempts to explain this shoulder by Si\,{\sc i} line blends seen in the
first order of diffraction (whereas the Ne\,{\sc viii} line is recorded in
the second order with the SUMER detector~A) were not successful
\citep{Dametal,Wil00}. The conclusion was that two
spectral components with a Doppler separation of 34~km~s$^{-1}$ were present
nearly symmetrically with respect to the rest wavelength, but the nature of
these components remained unclear.

The main purpose of this section and the next is to clarify
the situation and provide a physical explanation for the Ne\,{\sc viii}
profile in PCHs.

The observations re-analysed here have been presented by
\citet{Hasetal9,Dametal,Wil00}.
We consider the Ne\,{\sc viii} line profile in Fig.~5 of the latter paper and
apply multi-Gaussian fits on them.
This line is very weak in PCHs \citep[Fig.~7 of][]{Wil98} and,
in particular, in regions with high outflow speeds
\citep[Fig.~3 of][]{Wil00}, presumably IPRs.
We, therefore, assume that most of the radiation analysed in
Fig.~\ref{f.second_order}a stems from bright plumes and not from IPRs.

The results are shown separately in
Fig.~\ref{f.second_order}a and b for the PCH and
QS regions. They confirm that the QS
profile is of a near Gaussian shape. The PCH profile, however, is built up of
three components in the 2nd order spectrum: (1) A blue-shifted
component (Doppler shift: 19~km~s$^{-1}$) with a relative
contribution of $\approx 45$~\% to the total line radiance; (2) one with a
redshift of 15~km~s$^{-1}$ and a contribution of 35~\%; (3) a component with
a blueshift of $\approx 14$~km~s$^{-1}$ and a 15~\% contribution.
All Doppler shifts refer, of course, to line-of-sight components.
The outflow speeds along the magnetic field lines are approximately a factor
of 1.4 higher (see Sect.~\ref{s.outflow}). The weak 15~\% peak
will be attributed to the outflow in the IPRs with about 14~km~s$^{-1}$.

Although several first-order Si\,{\sc i}
lines blend the Ne\,{\sc viii} line, they do not produce the shoulder
(as mentioned above), and can be disregarded here. The red-shifted component
(the shaded area in Fig.~\ref{f.second_order}a)
is therefore difficult to understand and is an important topic of this
article.
An explanation will be presented in the next section based on a specific
plume model.


\section{Proposed plume model}
\label{s.Model}

An outflow velocity of $V_{\rm out} \approx 300$~km\,s$^{-1}$ of H$^0$
is reached in IPRs at $R \approx 3\,R_\odot$
\citep{Koh98}. We will approximate $V_\parallel$, the component
parallel to the magnetic field, by $V_{\rm out}$, and take the
transverse velocity, $V_\perp$,
also into account in defining the total velocity
%
\begin{equation}
V = \sqrt{V_\parallel^2 + V_\perp^2} \quad .
\label{velocity}
\end{equation}
The Ultraviolet Coronagraph Spectrometer (UVCS) observations on SOHO of
H\,{\sc i} Ly\,$\alpha$ line-width indicate that
$V_\perp$ of protons is of the order of 200~km\,s$^{-1}$ at $3\,R_\odot$, if
charge-exchange processes equalize the hydrogen and proton speeds. This
gives $V \approx 360$~km\,s$^{-1}$ and, consequently, most of the material
below $\approx 3\,R_\odot$ is still gravitationally bound to the Sun
(cf., also Fig.~\ref{f.escape}) as long as no post-acceleration is in operation.
Such an acceleration depends on waves generated by reconnection processes at
or near the footpoints of the funnel
\citep[cf., e.g.,][]{Ofm06}.

Given the fact that plumes exist on open magnetic field structures,
their geometries are not too different from the magnetic funnels described,
e.g. by \citet{Tuetal}, for CH regions in general.
The first question is: what are the conditions for a plume formation
compared to those for a \emph{normal} funnel?
The funnel activity consists of small-scale reconnection events close to
the TR and in the low corona.
This, in turn, creates heated plasma and waves that are
obviously capable of expanding the coronal plasma against gravity and
accelerating the fast SW to speeds of
$\approx 800$~km\,s$^{-1}$.
One answer could be that such an active
funnel at some stage ``burns out''. The alternative
answer that the activity in a funnel has not yet reached the
level required to produce the SW is less likely in view of observed BP/plume
evolution sequences \citep[cf.,][]{Wan98}.

Let us now consider how such a shutdown could happen. A narrow
funnel interacting with advected small loops will\,--\,in addition
to generating particle and wave energy\,--\,grow through reconnection.
However, not all of the advected loops will have the right
orientation for a successful interaction. These loops will accumulate around
the funnel and might eventually shield it from loops capable of creating
reconnection events. This configuration, if visualized in three
dimensions, resembles with that of a rosette, a
characteristic magnetic feature in the chromospheric network.

The scenario described will not lead to an abrupt shutdown of an active
funnel, but to a slow diminution of the reconnection activity, presumably
with the effect that plasma is injected into the funnel without enough
post-acceleration to leave the gravitational potential of the Sun. The
situation is now comparable to regions on the Sun with closed magnetic field
regions\,--\,the plasma density will increase and the FIP differentiation
would commence.
A coronal plume is formed, and at its base
a BP might be seen during this phase.

At a later stage, the energy input by reconnection will
decrease even more.
One could speculate that this is related to the fact that the cross-section of a
growing funnel base will increase faster than its circumference. The BP will
fade out, but the plume will not immediately collapse under the gravitational
pull of the Sun
as one might think, even if the thermal energy could be dumped at the base of
the plume, which is now assumed to be cool.
To show this, we will treat the plasma of the plume in a
single-particle approximation, justified by the low density of
$n_{\rm e} \leq 1 \times 10^8$~cm$^{-3}$ \citep[cf.,][]{Lie02}.
In a low $\beta$ regime of a magnetized plasma, the
protons will have a magnetic moment that can be written in the non-relativistic
case as
%
\begin{equation}
\mu_{\rm p} = \frac{[W - m_{\rm p}\,U(s)]}{B(s)}\,{\rm sin}^2 \alpha(s) ~,
\label{Eqmom}
\end{equation}
where $W$ is the total proton energy, $m_{\rm p}$ the proton mass,
$U(s)$ the gravitational potential with $s$ a spatial parameter along the
field direction,
$B(s)$ the magnetic field and the pitch angle, $\alpha$, defined by
%
\begin{equation}
\alpha = {\rm arc cos} \, \frac{V_\parallel}{V} ~.
\label{Eqpit}
\end{equation}
The magnetic moment will be a constant of the motion, the first adiabatic
invariant, as long as Coulomb collisions and
wave-particle interactions can be neglected.
This concept was first formulated by Alfv\'en, details of which are described,
for instance, by \citet{Roe70} for applications in the magnetosphere of the
Earth.

If we consider a plume at one of the solar poles,
the gravitational potential is
%
\begin{equation}
U(s) = U(R) = -\frac{G_{\rm N}\,M_\odot}{R} ~ .
\label{Eqpot}
\end{equation}
Assuming a magnetic pole at $R = 0.56~R_\odot$ \citep[cf.,][]{Sai65},
we find a variation of the field in a PCH as
%
\begin{equation}
B(R) \approx B_0\,\left(\frac{R_\odot}{R}\right)^{3.6} ~ ,
\label{Eqfie}
\end{equation}
where $B_0$ is the field at the pole.
In such a geometry, most of the plasma is trapped. With the
assumptions made, it is immediately clear that it cannot leave the Sun at the
upper end of the flux tube. The continuation of the plume flux tube
will thus be more or less void of plasma \citep[cf.,][]{Wiletal}.
At the sunward side only particles
within the loss cone will be lost, the others will be mirrored.
\citet{Hul50} demonstrated that the Lorentz force will not influence the
hydrostatic equilibrium, but in our configuration it will confine the plume
plasma during the radiative cooling phase. During this phase, we would
expect plasma flows up and down the plume field lines and suggest that
the blue- and red-shifted strong components in Fig.~\ref{f.second_order}a
correspond to these flows, if seen along the line of sight.


\section{Conclusion}
\label{s.concl}

The different elemental compositions of plumes and IPRs strongly suggest that
plumes in contrast to IPRs provide some kind of containment for the solar
plasma for a period of days, in which the FIP effect can operate.

The continuation of the plume flux tube may be more or less void of plasma
\citep[cf.,][]{Wiletal}. At the sunward side, only particles within the
loss cone will be lost, the others will be mirrored. \citet{Hul50}
demonstrated that the Lorentz force will not influence the hydrostatic
equilibrium. In our model, it will confine the plume plasma during the
radiative cooling phase. During this phase, we would expect that plasma flows
up and down the plume field lines and that the strong blue- and red-shifted
components of the neon line correspond to these flows.

\vspace{0.5cm}


\begin{acknowledgements}
We thank the Max-Planck-Institut f\"ur Sonnensystemforschung for
administrative support
and an anonymous referee for constructive comments.
The SUMER instrument and its operation are financed by the
Deutsches Zentrum f\"ur Luft- und Raumfahrt (DLR), the Centre National
d'Etudes Spatiales (CNES), the National Aeronautics and Space Administration
(NASA), and the European Space Agency's (ESA) PRODEX programme
(Swiss contribution). The instrument is part of ESA's and NASA's SOHO
mission.
This research has made extensive use of the Astrophysics Data System (ADS).
\end{acknowledgements}



\begin{thebibliography}{00}
%
\bibitem[Antonucci et al.(2004)]{Antetal}
Antonucci, E., Dodero, M.A., Giordano, S., Krishnakumar, V., \&  Noci, G.
2004,
\aap, 416, 749
%
\bibitem[Banaszkiewicz et al.(1998)]{Banetal}
Banaszkiewicz, M., Axford,  W.I., \&  McKenzie, J.F. 1998,
\aap, 337, 940
%
\bibitem[Brekke et al.(1997)]{Breetal}
Brekke, P., Hassler, D.M., \& Wilhelm, K. 1997,
\solphys, 175, 349
%
\bibitem[Chae et al.(1997)]{Chaetal}
Chae, J., Yun, H.S., \&  Poland, A.I. 1997,
\apjs, 114, 151
%
\bibitem[Dammasch et al.(1999)]{Dametal}
Dammasch, I.E., Wilhelm, K., Curdt, W., \& Hassler, D.M. 1999,
\aap, 346, 285
%
\bibitem[DeForest et al.(1997)]{DeFetal}
DeForest, C.E., Hoeksema, J.T., Gurman, J.B., et al. 1997,
\solphys, 175, 393
%
\bibitem[DeForest et al.(2001a)]{DeF01a}
DeForest, C.E., Lamy, P.L., \&  Llebaria, A. 2001a,
\apj, 560, 490
%
\bibitem[DeForest et al.(2001b)]{DeF01b}
DeForest, C.E., Plunkett, S.P., \& Andrews, M.D. 2001b,
\apj, 546, 569
%
\bibitem[Delaboudini\`ere et al.(1995)]{Boudine}
Delaboudini\`ere, J.-P., Artzner, G.E., Brunaud, J., et al. 1995,
\solphys, 162, 291
%
\bibitem[Doschek et al.(1976)]{Dosetal}
Doschek, G.A., Feldman, U., \& Bohlin, J.D. 1976,
\apj, 205, L177
%
\bibitem[Doschek et al.(1998)]{Dosetal8}
Doschek, G.A., Laming, J.M., Feldman, U., et al. 1998,
\apj, 504, 573
%
\bibitem[Fawcett et al.(1961)]{Fawetal}
Fawcett, B.C., Jones, B.B., \& Wilson, R. 1961,
Proc. Phys. Soc., 78, 1223
%
\bibitem[Feldman(1998)]{Fel98}
Feldman, U. 1998,
\ssr, 85, 227
%
\bibitem[Gabriel(1976)]{Gab76}
Gabriel, A.H. 1976,
Phil. Trans. R. Soc. London, Ser. A, 281, 339
%
\bibitem[Gabriel et al.(2003)]{Gab03}
Gabriel, A.H., Bely-Dubau, F., \& Lemaire, P. 2003,
\apj, 589, 623
%
\bibitem[Gabriel et al.(2009)]{Gab09}
Gabriel, A., Bely-Dubau, F., Tison, E., \& Wilhelm, K. 2009,
\apj, 700, 551
%
\bibitem[Geiss et al.(1995)]{Geietal}
Geiss, J., Gloeckler, G., \& von Steiger, R. 1995,
\ssr, 72, 49
%
\bibitem[Hassler et al.(1991)]{Hasetal1}
Hassler, D.M., Rottman, G.J., \& Orrall, F.Q. 1991,
\apj, 372, 710
%
\bibitem[Hassler et al.(1999)]{Hasetal9}
Hassler, D.M., Dammasch, I.E. Lemaire, P., et al. 1999,
Science, 283, 810
%
\bibitem[Heber et al.(2013)]{Hebetal}
Heber, V.S., McKeegan, K.D., Bochsler, P., et al. 2013,
Lunar Planet. Inst., 44, 3028
%
\bibitem[van de Hulst(1950)]{Hul50}
van de Hulst, H.C. 1950,
Bull. Astron. Inst. Netherlands, 11, 150
%
\bibitem[Ito et al.(2010)]{Itoetal}
Ito, H., Tsuneta, S., Shiota, D., Tokumaru, M., \& Fujiki, K. 2010,
\apj, 719, 131
%
\bibitem[Kohl et al.(1998)]{Koh98}
Kohl, J.L., Noci, G., Antonucci, E., et al. 1998,
\apj, 501, L127
%
\bibitem[Landi(2008)]{Lan08}
Landi, E. 2008,
\apj, 685, 1270
%
\bibitem[Lie-Svendsen et al.(2002)]{Lie02}
Lie-Svendsen, {\O}., Hansteen, V.H., Leer, E., \&  Holzer, T.E. 2002,
\apj, 566, 562
%
\bibitem[McComas et al.(2000)]{McCetal}
McComas, D.J., Barraclough, B.L., Funsten, H.O., et al. 2000,
J. Geophys. Res., 105, 10\,419
%
\bibitem[Neugebauer et al.(1995)]{Neuetal}
Neugebauer, M., Goldstein, B.E., McComas, D.J., Suess, S.T., \&  Balogh, A.
1995,
J. Geophys. Res. 100, 23\,389
%
\bibitem[Newkirk and Harvey(1968)]{NewHar}
Newkirk, Jr., G., \& Harvey, J. 1968,
\solphys, 3, 321
%
\bibitem[Ofman(2006)]{Ofm06}
Ofman, L. 2006,
Adv. Space Res., 38, 64
%
\bibitem[Pasachoff et al.(2009)]{Pasetal}
Pasachoff, J.M., Ru\v{s}in, V., Druckm\"uller, M., et al. 2009
\apj, 702, 1297
%
\bibitem[de Patoul et al.(2013)]{Patetal}
de Patoul, J., Inhester, B., Feng, L., \& Wiegelmann, T. 2013,
\solphys, 283, 207
%
\bibitem[Peter and Judge(1999)]{PetJud}
Peter H., \& Judge, P.G. 1999,
\apj, 522, 1148
%
\bibitem[Poletto et al.(1996)]{Poletal}
Poletto, G., Parenti, S., Noci, G., et al. 1996,
\aap, 316, 374
%
\bibitem[Raouafi \& Stenborg(2014)]{RaoSte}
Raouafi, N.-E., \& Stenborg, G. 2014,
\apj, 787, 118
%
\bibitem[Raouafi et al.(2008)]{Raoetal}
Raouafi, N.-E., Petrie, G.J.D., Norton, A.A., Henney, C.J.,
\&  Solanki, S. K. 2008,
\apj, 682, L137
%
\bibitem[Reisenfeld et al.(1999)]{Reietal}
Reisenfeld, D.B., Mc Comas, D.J., \& Steinberg, J.T. 1999,
\grl, 26, 1805
%
\bibitem[Roederer(1970)]{Roe70}
Roederer, J.G. 1970,
Dynamics of geomagnetically trapped radiation,
(Springer-Verlag, Berlin, Heidelberg, New York)
%
%
\bibitem[Saito(1965)]{Sai65}
Saito, K. 1965,
Publ. Astron. Soc. Japan, 17, 1
%
\bibitem[Sheeley et al.(1997)]{Sheetal}
Sheeley, Jr., N.R., Wang, Y.-M., Hawley, S.H., et al. 1997,
\apj, 484, 472
%
\bibitem[Sheminova and Solanki(1999)]{SheSol}
Sheminova, V.A., \& Solanki, S.K. 1999,
\aap, 351, 701
%
\bibitem[von Steiger et al.(1999)]{Steetal}
von Steiger, R., Fisk, L.A., Gloeckler, G., Schwadron, N.A.,
\& Zurbuchen, T.H. 1999,
AIP Conf. Proc., 471, 143
%
\bibitem[Thieme et al.(1990)]{Thi90}
Thieme, K.M., Marsch, E., \& Schwenn, R. 1990,
Ann. Geophys. 8, 713
%
\bibitem[Tu et al.(2005)]{Tuetal}
Tu, C.-Y., Zhou, C., Marsch, E., et al. 2005,
Science, 308, 519
%
\bibitem[Wang(1998)]{Wan98}
Wang, Y.-M. 1998,
\apj, 501, L145
%
\bibitem[Widing and Feldman(1992)]{WidFel9}
Widing, K.G., \& Feldman, U. 1992,
\apj, 392, 715
%
\bibitem[Widing and Feldman(2001)]{WidFel2}
Widing, K.G., \& Feldman, U. 2001,
\apj, 555, 426
%
\bibitem[Wilhelm(2006)]{Wil06}
Wilhelm, K. 2006,
\aap, 455, 697
%
\bibitem[Wilhelm and Bodmer(1998)]{WilBod}
Wilhelm, K., \& Bodmer, R. 1998,
\ssr, 85, 371
%
\bibitem[Wilhelm et al.(1995)]{Wil95}
Wilhelm, K., Curdt, W., E. Marsch, E., et al. 1995,
\solphys, 162, 189
%
\bibitem[Wilhelm et al.(1998)]{Wil98}
Wilhelm, K., Lemaire, P.,  Dammasch, I.E., et al. 1998,
\aap, 334, 685
%
\bibitem[Wilhelm et al.(2000)]{Wil00}
Wilhelm, K., Dammasch, I.E., Marsch, E., \& Hassler, D.M. 2000,
\aap, 353, 749
%
\bibitem[Wilhelm et al.(2002)]{Wil02}
Wilhelm, K., Inhester, B., \& Newmark, J.S. 2002,
\aap, 382, 328
%
\bibitem[Wilhelm et al.(2011)]{Wiletal}
Wilhelm, K., Abbo, L., Auch\`ere, F., et al. 2011,
\aapr, 19, 35
%
\bibitem[Woch et al.(1997)]{Wocetal}
Woch, J., Axford, W.I., Mall, U., et al. 1997,
\grl, 24, 2885
%
\bibitem[Young et al.(1999)]{Youetal}
Young, P.R., Klimchuk, J.A., \&  Mason, H.E. 1999,
\aap, 350, 286
%
\end{thebibliography}
\end{document}